\documentclass[showpacs,aps,graphicx]{revtex4}
\usepackage{graphicx}

\begin{document}
\title{The heralded amplification for the single-photon multi-mode W state of the time-bin qubit}

\author{Lan Zhou$^{1,2}$, Yu-Bo Sheng,$^{2,3}$\footnote{Email address:
shengyb@njupt.edu.cn}  }
\address{
 $^1$ College of Mathematics \& Physics, Nanjing University of Posts and Telecommunications, Nanjing,
210003, China\\
 $^2$Key Lab of Broadband Wireless Communication and Sensor Network
 Technology, Nanjing University of Posts and Telecommunications, Ministry of
 Education, Nanjing, 210003, China\\
$^3$Institute of Signal Processing  Transmission, Nanjing
University of Posts and Telecommunications, Nanjing, 210003,  China\\}

\begin{abstract}
We put forward an effective amplification protocol for protecting the single-photon multi-mode W state of the time-bin qubit. The protocol only relies on linear optical elements, such as the $50:50$ beam splitters, variable beam splitters with the transmission of $t$ and the polarizing beam splitters. Only one pair of the single-photon multi-mode W state and some auxiliary single photons are required, and the fidelity of the single-photon multi-mode W state can be increased under $t<\frac{1}{2}$. The encoded time-bin information can be perfectly contained. Our protocol is quite simple and economical, and it can be realized under current experimental condition. Based on above features, it may be useful in current and future quantum information processing.
\end{abstract}
 \maketitle

\section{Introduction}
Quantum entanglement is an importance source in quantum communications \cite{teleportation,teleportation1,Ekert91,qkd1,qkd2,QSS,QSDC1,QSDC2}. For example, in the fields of quantum teleportation \cite{teleportation,teleportation1}, quantum key distribution (QKD) \cite{Ekert91,qkd1,qkd2}, quantum state sharing (QSS) \cite{QSS}, quantum secure direct communication (QSDC) \cite{QSDC1,QSDC2}, the parties all resort to the entanglement to set up the quantum channel. Entanglement is also essential in quantum computations \cite{dengcomputation4,dengcomputation5,wangcomputation}. In various entanglement forms, the single photon entanglement (SPE) is the simplest but important one. The typical form of the SPE is
\begin{eqnarray}
|SPE\rangle=\frac{1}{\sqrt{2}}(|10\rangle_{AB}+|01\rangle_{AB}),\label{SPE}
\end{eqnarray}
where A and B are two different spatial modes,  $|0\rangle$ and $|1\rangle$ mean no photon and one photon, respectively. The
SPE between two modes has been
proved to be a valuable resource for cryptography \cite{cryptography1,cryptography2},
state engineering \cite{engineering}, and tomography of states and
operations \cite{tomography1,tomography2}. In the past few years, researchers have extended the two-mode SPE in Eq. (\ref{SPE}) to the N-mode W state with the form
\begin{eqnarray}
|W\rangle=\frac{1}{\sqrt{N}}(|100\cdots0\rangle+|010\cdots0\rangle+\cdots+|00\cdots1\rangle),
\end{eqnarray}
where the single photon is superpositioned in N spatial modes in different spatial locations ($N>2$). The single-photon W state displays an effective all-versus nothing
nonlocality as the number of N delocalizations of
the single particle goes up \cite{dolocalization}, and it is robust to decoherence in the noisy
environment \cite{robust1,robust2,robust3}. In 2012, Gottesman \emph{et al.} proposed a protocol for
building an interferometric telescope based on the single-photon multi-mode W state \cite{SPE3}. The protocol has the potential to eliminate the
baseline length limit, and allows the interferometers
with arbitrarily long baselines in principle.

In the paper, we focus on the single-photon W state of the time-bin entangled qubit. The time-bin entanglement is a coherent superposition of single photon state in two or more different temporal modes. Suppose a single-photon pulse enters two channels, where channel length difference (long-short) is much longer than the pulse duration. The output state consists of two well separated pulses with different times of arrival, and the quantum information is encoded in the arrival
time of the photon, which can be written as $|Long\rangle$ ($|L\rangle$) and $|Short\rangle$ ($|S\rangle$), respectively. The time-bin entanglement is a robust form of optical quantum
information, especially for the transmission in optical fibers. It has been widely used in the long-distance entanglement distribution \cite{distribution1,distribution2,distribution3}. For example, in
2002, Thew \emph{et al.} demonstrated the robust time-bin qubits for distributed
quantum communication over $11$ $km$ \cite{distribution1}. Later,
Marcikic \emph{et al.} experimentally realized the distribution of
time-bin entangled qubits over $50$ $km$ of optical fibers \cite{distribution2}. Recently, Inagaki \emph{et al.} has realized the distribution of time-bin entangled photons over $300$ $km$ of optical fiber \cite{distribution3}. By distributing the time-bin entangled qubit to N spatial modes in different spatial locations, we can create a single-photon N-mode W state of the time-bin qubit.

Unfortunately, the photon may be completely lost in the distribution process. The photon loss will make the pure state mix with the vacuum state with some probability. In this way, the entanglement between two distant locations decreases exponentially with the length of the transmission channel. The noiseless linear amplification (NLA) is an effective method to overcome the exponential
fidelity decay \cite{NLA1,NLA2,NLA3,NLA4,NLA5,NLA6,NLA7,NLA8,NLA9,NLA10,NLA11,NLA12,NLA13,NLA14,NLA15,nonlinear}. In 2012, Zhang \emph{et al.} proposed the NLA to protect
the two-mode SPE from photon loss \cite{NLA8}. In 2015, Sheng \emph{et al.} successfully protect the SPE with the experimentally available parametric down-conversion (SPDC) source \cite{NLA15}. In the same year, Zhou and Sheng proposed a recyclable amplification protocol for the SPE with the weak cross-Kerr nonlinearity. By repeating this amplification, they can effectively increase the fidelity of the output state \cite{nonlinear}. Recently, a linear heralded qubit amplification protocol for the time-bin qubit and polarization qubit is proposed by Bruno \emph{et al.} \cite{timebin}, which can be realized under current experimental conditions. Inspired by this protocol, our group put forward a simple and effective NLA protocol for the SPE of the time-bin qubit \cite{zhoutimebin}. Here, in our paper, we extend this NLA protocol in Ref. \cite{zhoutimebin} to the single-photon multi-mode W state of the time-bin entangled qubit. We will show that our protocol can effectively increase the fidelity of the single-photon multi-mode W state. Our protocol will be useful in the current and future quantum communication field.

 The paper is organized as follows, in Sec. 2, we explain our amplification protocol for the single-photon three-mode W state of the time-bin qubit, in Sec. 3, we extend the amplication protocol to the arbitrary single-photon multi-mode W state, in Sec. 4, we calculate the amplification factor, and the success probability of our protocol, and make a discussion and summary.

\section{The heralded amplification for the single-photon three-mode W state of the time-bin qubit}
\begin{figure}[!h]
\begin{center}
\includegraphics[width=10cm,angle=0]{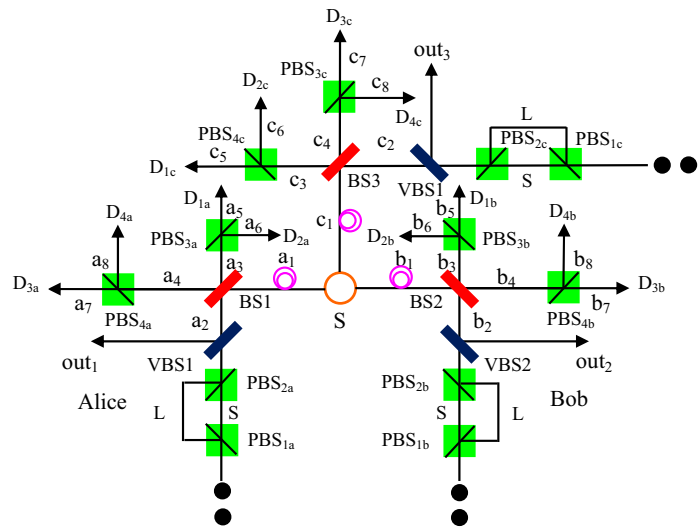}
\caption{The schematic principle of the amplification protocol for the single-photon three-mode W state of the time-bin qubit. Here, BS represents the 50:50 beam splitter, VBS represents the variable beam splitter with the transmission of $t$, and PBS means the polarizing beam splitter. }
\end{center}
\end{figure}

In this section, we will explain our amplification protocol for the single-photon three-mode W state of the time-bin qubit. Suppose a time-bin qubit has the form of
\begin{eqnarray}
|\psi\rangle=\alpha|S_{H}\rangle+\beta|L_{V}\rangle,\label{timebin}
\end{eqnarray}
where $|S\rangle$ ($|L\rangle$) means the $short$ ($long$) time-bin. The subscript $H$ or $V$ indicates that the photon is in horizontal polarization or vertical polarization. Here, the polarization is used to label and switch the path of the photon. $\alpha$ and $\beta$ are the entanglement coefficients, where $|\alpha|^{2}+|\beta|^{2}=1$. The time-bin qubit $|\psi\rangle$ is distributed to three parties, say, Alice (A), Bob (B) and Charlie (C), which can create a single-photon three-mode W state as
\begin{eqnarray}
|\Phi\rangle_{ABC}=\frac{1}{\sqrt{3}}(|\psi\rangle_{a1}|0\rangle_{b1}|0\rangle_{c1}+|0\rangle_{a1}|\psi\rangle_{b1}|0\rangle_{c1}
+|0\rangle_{a1}|0\rangle_{b1}|\psi\rangle_{c1}).\label{max}
\end{eqnarray}

Due to the environmental noise in the distribution channel, the state in Eq. (\ref{max}) may be mixed with the vacuum state. We suppose the single-photon can be completely lost with the probability of $1-\eta$. In this way, the three parties share a mixed state as
 \begin{eqnarray}
\rho_{in}=\eta|\Phi\rangle_{ABC}\langle\Phi|+(1-\eta)|vac\rangle\langle vac|.
\end{eqnarray}
The aim of the heralded amplification protocol is to increase the fidelity $\eta$.

The schematic principle of the protocol is shown in Fig. 1. For realizing the amplification, each of the three parties needs to prepare two auxiliary single photons, one in $|H\rangle$ and the other one in $|V\rangle$. All the parties make the auxiliary photons in their hands pass through two polarizing beam splitters (PBSs), say $PBS_{1a}PBS_{2a}$, $PBS_{1b}PBS_{2b}$, and $PBS_{1c}PBS_{2c}$, respectively. The PBS can totally transmit the photon in $|H\rangle$ but totally reflect the photon in $|V\rangle$. Due to the path-length difference of the photons in $|H\rangle$ and $|V\rangle$, each of they can create an auxiliary photon state as $|\varphi\rangle=|S_{H}\rangle\otimes|L_{V}\rangle$.

Then, each of the parties  makes the auxiliary photons in his or her hand pass through one variable beam splitter (VBS), say VBS1, VBS2, and VBS3. All the three VBSs have the same transmission of $t$. After the VBSs, they can obtain
\begin{eqnarray}
|\varphi_{ABC}\rangle&=&(\sqrt{t}|S_{H}0\rangle_{a2out1}+\sqrt{1-t}|0S_{H}\rangle_{a2out1})\otimes(\sqrt{t}|L_{V}0\rangle_{a2out1}+\sqrt{1-t}|0L_{V}\rangle_{a2out1})\nonumber\\
&\otimes&(\sqrt{t}|S_{H}0\rangle_{b2out2}+\sqrt{1-t}|0S_{H}\rangle_{b2out2})\otimes(\sqrt{t}|L_{V}0\rangle_{b2out2}+\sqrt{1-t}|0L_{V}\rangle_{b2out2})\nonumber\\
&\otimes&(\sqrt{t}|S_{H}0\rangle_{c2out3}+\sqrt{1-t}|0S_{H}\rangle_{c2out3})\otimes(\sqrt{t}|L_{V}0\rangle_{c2out3}+\sqrt{1-t}|0L_{V}\rangle_{c2out3})\nonumber\\
&=&[t|S_{H}L_{V}\rangle_{a2}+(1-t)|S_{H}L_{V}\rangle_{out1}+\sqrt{t(1-t)}(|S_{H}L_{V}\rangle_{a2out1}+|L_{V}S_{H}\rangle_{a2out1})]\nonumber\\
&\otimes&[t|S_{H}L_{V}\rangle_{b2}+(1-t)|S_{H}L_{V}\rangle_{out2}+\sqrt{t(1-t)}(|S_{H}L_{V}\rangle_{b2out2}+|L_{V}S_{H}\rangle_{b2out2})]\nonumber\\
&\otimes&[t|S_{H}L_{V}\rangle_{c2}+(1-t)|S_{H}L_{V}\rangle_{out3}+\sqrt{t(1-t)}(|S_{H}L_{V}\rangle_{c2out3}+|L_{V}S_{H}\rangle_{c2out3})].
\end{eqnarray}
 In this way, the whole state $\rho_{in}\otimes|\varphi_{ABC}\rangle$ can be in the state of $|\Phi\rangle_{ABC}\otimes|\varphi_{ABC}\rangle$ with the probability of $\eta$, or in the state of $|vac\rangle\otimes|\varphi_{ABC}\rangle$ with the probability of $1-\eta$. We first discuss the case of $|\Phi\rangle_{ABC}\otimes|\varphi_{ABC}\rangle$, which can be written as
\begin{eqnarray}
|\Phi\rangle_{ABC}\otimes|\varphi_{ABC}\rangle&=&\frac{1}{\sqrt{3}}[(\alpha|S_{H}\rangle_{a1}+\beta|L_{V}\rangle_{a1})|0\rangle_{b1}|0\rangle_{c1}
+|0\rangle_{a1}(\alpha|S_{H}\rangle_{b1}+\beta|L_{V}\rangle_{b1})|0\rangle_{c1}\nonumber\\&&+
|0\rangle_{a1}|0\rangle_{b1}(\alpha|S_{H}\rangle_{c1}+\beta|L_{V}\rangle_{c1})]\nonumber\\
&\otimes&[t|S_{H}L_{V}\rangle_{a2}+(1-t)|S_{H}L_{V}\rangle_{out1}+\sqrt{t(1-t)}(|S_{H}L_{V}\rangle_{a2out1}+|L_{V}S_{H}\rangle_{a2out1})]\nonumber\\
&\otimes&[t|S_{H}L_{V}\rangle_{b2}+(1-t)|S_{H}L_{V}\rangle_{out2}+\sqrt{t(1-t)}(|S_{H}L_{V}\rangle_{b2out2}+|L_{V}S_{H}\rangle_{b2out2})]\nonumber\\
&\otimes&[t|S_{H}L_{V}\rangle_{c2}+(1-t)|S_{H}L_{V}\rangle_{out3}+\sqrt{t(1-t)}(|S_{H}L_{V}\rangle_{c2out3}+|L_{V}S_{H}\rangle_{c2out3})].
\label{whole1}
\end{eqnarray}

  The three parties make the photons in the $a_{1}a_{2}$, $b_{1}b_{2}$, and $c_{1}c_{2}$ modes pass through the $50:50$ beam splitters (BSs), here named BS1, BS2 and BS3, respectively. The BSs can make
\begin{eqnarray}
|1\rangle_{a1}&=&\frac{1}{\sqrt{2}}(|1\rangle_{a3}+|1\rangle_{a4}),\quad |1\rangle_{a2}=\frac{1}{\sqrt{2}}(|1\rangle_{a3}-|1\rangle_{a4}),\nonumber\\
|1\rangle_{b1}&=&\frac{1}{\sqrt{2}}(|1\rangle_{b3}+|1\rangle_{b4}),\quad |1\rangle_{b2}=\frac{1}{\sqrt{2}}(|1\rangle_{b3}-|1\rangle_{b4}).\nonumber\\
|1\rangle_{c1}&=&\frac{1}{\sqrt{2}}(|1\rangle_{c3}+|1\rangle_{c4}),\quad |1\rangle_{c2}=\frac{1}{\sqrt{2}}(|1\rangle_{c3}-|1\rangle_{c4}).
\end{eqnarray}
After that, the state in Eq. (\ref{whole1}) will evolve to
\begin{eqnarray}
|\Phi\rangle_{ABC}\otimes|\varphi_{ABC}\rangle&\rightarrow&\frac{1}{\sqrt{6}}\{(\alpha|S_{H}\rangle_{a3}+\alpha|S_{H}\rangle_{a4}+\beta|L_{V}\rangle_{a3}+\beta|L_{V}\rangle_{a4})\nonumber\\
&\otimes&[\frac{t}{2}(|S_{H}L_{V}\rangle_{a3}-|S_{H}L_{V}\rangle_{a3a4}-|L_{V}S_{H}\rangle_{a3a4}+|S_{H}L_{V}\rangle_{a4})+(1-t)|S_{H}L_{V}\rangle_{out1}\nonumber\\
&+&\frac{\sqrt{t(1-t)}}{\sqrt{2}}(|S_{H}L_{V}\rangle_{a3out1}-|S_{H}L_{V}\rangle_{a4out1}-|L_{V}S_{H}\rangle_{a3out1}-|L_{V}S_{H}\rangle_{a4out1})]\nonumber\\
&\otimes&[\frac{t}{2}(|S_{H}L_{V}\rangle_{b3}-|S_{H}L_{V}\rangle_{b3b4}-|L_{V}S_{H}\rangle_{b3b4}+|S_{H}L_{V}\rangle_{b4})
+(1-t)|S_{H}L_{V}\rangle_{out2}\nonumber\\
&+&\frac{\sqrt{t(1-t)}}{\sqrt{2}}(|S_{H}L_{V}\rangle_{b3out2}-|S_{H}L_{V}\rangle_{b4out2}-|L_{V}S_{H}\rangle_{b3out2}-|L_{V}S_{H}\rangle_{b4out2})]\nonumber\\
&\otimes&[\frac{t}{2}(|S_{H}L_{V}\rangle_{c3}-|S_{H}L_{V}\rangle_{c3c4}-|L_{V}S_{H}\rangle_{c3c4}+|S_{H}L_{V}\rangle_{c4})+(1-t)|S_{H}L_{V}\rangle_{out3}\nonumber\\
&+&\frac{\sqrt{t(1-t)}}{\sqrt{2}}(|S_{H}L_{V}\rangle_{c3out3}-|S_{H}L_{V}\rangle_{c4out3}-|L_{V}S_{H}\rangle_{c3out3}-|L_{V}S_{H}\rangle_{c4out3})]\nonumber\\
&+&(\alpha|S_{H}\rangle_{b3}+\alpha|S_{H}\rangle_{b4}+\beta|L_{V}\rangle_{b3}+\beta|L_{V}\rangle_{b4})\nonumber\\
&\otimes&[\frac{t}{2}(|S_{H}L_{V}\rangle_{a3}-|S_{H}L_{V}\rangle_{a3a4}-|L_{V}S_{H}\rangle_{a3a4}+|S_{H}L_{V}\rangle_{a4})
+(1-t)|S_{H}L_{V}\rangle_{out1}\nonumber\\
&+&\frac{\sqrt{t(1-t)}}{\sqrt{2}}(|S_{H}L_{V}\rangle_{a3out1}-|S_{H}L_{V}\rangle_{a4out1}-|L_{V}S_{H}\rangle_{a3out1}-|L_{V}S_{H}\rangle_{a4out1})]\nonumber\\
&\otimes&[\frac{t}{2}(|S_{H}L_{V}\rangle_{b3}-|S_{H}L_{V}\rangle_{b3b4}-|L_{V}S_{H}\rangle_{b3b4}+|S_{H}L_{V}\rangle_{b4})
+(1-t)|S_{H}L_{V}\rangle_{out2}\nonumber\\
&+&\frac{\sqrt{t(1-t)}}{\sqrt{2}}(|S_{H}L_{V}\rangle_{b3out2}-|S_{H}L_{V}\rangle_{b4out2}-|L_{V}S_{H}\rangle_{b3out2}-|L_{V}S_{H}\rangle_{b4out2})]\nonumber\\
&\otimes&[\frac{t}{2}(|S_{H}L_{V}\rangle_{c3}-|S_{H}L_{V}\rangle_{c3c4}-|L_{V}S_{H}\rangle_{c3c4}+|S_{H}L_{V}\rangle_{c4})+(1-t)|S_{H}L_{V}\rangle_{out3}\nonumber\\
&+&\frac{\sqrt{t(1-t)}}{\sqrt{2}}(|S_{H}L_{V}\rangle_{c3out3}-|S_{H}L_{V}\rangle_{c4out3}-|L_{V}S_{H}\rangle_{c3out3}-|L_{V}S_{H}\rangle_{c4out3})]\nonumber\\
&+&(\alpha|S_{H}\rangle_{c3}+\alpha|S_{H}\rangle_{c4}+\beta|L_{V}\rangle_{c3}+\beta|L_{V}\rangle_{c4})\nonumber\\
&\otimes&[\frac{t}{2}(|S_{H}L_{V}\rangle_{a3}-|S_{H}L_{V}\rangle_{a3a4}-|L_{V}S_{H}\rangle_{a3a4}+|S_{H}L_{V}\rangle_{a4})
+(1-t)|S_{H}L_{V}\rangle_{out1}\nonumber\\
&+&\frac{\sqrt{t(1-t)}}{\sqrt{2}}(|S_{H}L_{V}\rangle_{a3out1}-|S_{H}L_{V}\rangle_{a4out1}-|L_{V}S_{H}\rangle_{a3out1}-|L_{V}S_{H}\rangle_{a4out1})]\nonumber\\
&\otimes&[\frac{t}{2}(|S_{H}L_{V}\rangle_{b3}-|S_{H}L_{V}\rangle_{b3b4}-|L_{V}S_{H}\rangle_{b3b4}+|S_{H}L_{V}\rangle_{b4})
+(1-t)|S_{H}L_{V}\rangle_{out2}\nonumber\\
&+&\frac{\sqrt{t(1-t)}}{\sqrt{2}}(|S_{H}L_{V}\rangle_{b3out2}-|S_{H}L_{V}\rangle_{b4out2}-|L_{V}S_{H}\rangle_{b3out2}-|L_{V}S_{H}\rangle_{b4out2})]\nonumber\\
&\otimes&[\frac{t}{2}(|S_{H}L_{V}\rangle_{c3}-|S_{H}L_{V}\rangle_{c3c4}-|L_{V}S_{H}\rangle_{c3c4}+|S_{H}L_{V}\rangle_{c4})+(1-t)|S_{H}L_{V}\rangle_{out3}\nonumber\\
&+&\frac{\sqrt{t(1-t)}}{\sqrt{2}}(|S_{H}L_{V}\rangle_{c3out3}-|S_{H}L_{V}\rangle_{c4out3}-|L_{V}S_{H}\rangle_{c3out3}-|L_{V}S_{H}\rangle_{c4out3})]
\}.\label{whole2}
\end{eqnarray}

 Next, the parties make a Bell-state measurement (BSM) for the output photons. In detail, they make the photons in the $a_{3}a_{4}$, $b_{3}b_{4}$, and $c_{3}c_{4}$ modes pass through $PBS_{3a}PBS_{4a}$, $PBS_{3b}PBS_{4b}$, and $PBS_{3c}PBS_{4c}$, respectively, which can make
\begin{eqnarray}
|S_{H}\rangle_{i3}&\rightarrow&|S_{H}\rangle_{i5},\quad |L_{V}\rangle_{i3}\rightarrow|L_{V}\rangle_{i6},\nonumber\\
|S_{H}\rangle_{i4}&\rightarrow&|S_{H}\rangle_{i7},\quad |L_{V}\rangle_{i4}\rightarrow|L_{V}\rangle_{i8},
\end{eqnarray}
where $i=a, b, c$.

After the PBSs, the state in Eq. (\ref{whole2}) will evolve to
\begin{eqnarray}
|\Phi\rangle_{ABC}\otimes|\varphi_{ABC}\rangle&\rightarrow&\frac{1}{\sqrt{6}}\{(\alpha|S_{H}\rangle_{a5}+\alpha|S_{H}\rangle_{a7}+\beta|L_{V}\rangle_{a6}+\beta|L_{V}\rangle_{a8})\nonumber\\
&\otimes&[\frac{t}{2}(|S_{H}\rangle_{a5}|L_{V}\rangle_{a6}-|S_{H}\rangle_{a5}|L_{V}\rangle_{a8}-|L_{V}\rangle_{a6}|S_{H}\rangle_{a7}+|S_{H}\rangle_{a7}|L_{V}\rangle_{a8})+(1-t)|S_{H}L_{V}\rangle_{out1}\nonumber\\
&+&\frac{\sqrt{t(1-t)}}{\sqrt{2}}(|S_{H}\rangle_{a5}|L_{V}\rangle_{out1}-|S_{H}\rangle_{a7}|L_{V}\rangle_{out1}-|L_{V}\rangle_{a6}|S_{H}\rangle_{out1}-|L_{V}\rangle_{a8}|S_{H}\rangle_{out1})]\nonumber\\
&\otimes&[\frac{t}{2}(|S_{H}\rangle_{b5}|L_{V}\rangle_{b6}-|S_{H}\rangle_{b5}|L_{V}\rangle_{b8}-|L_{V}\rangle_{b6}|S_{H}\rangle_{b7}+|S_{H}\rangle_{b7}|L_{V}\rangle_{b8})+(1-t)|S_{H}L_{V}\rangle_{out2}\nonumber\\
&+&\frac{\sqrt{t(1-t)}}{\sqrt{2}}(|S_{H}\rangle_{b5}|L_{V}\rangle_{out2}-|S_{H}\rangle_{b7}|L_{V}\rangle_{out2}-|L_{V}\rangle_{b6}|S_{H}\rangle_{out2}-|L_{V}\rangle_{b8}|S_{H}\rangle_{out2})]\nonumber\\
&\otimes&[\frac{t}{2}(|S_{H}\rangle_{c5}|L_{V}\rangle_{c6}-|S_{H}\rangle_{c5}|L_{V}\rangle_{c8}-|L_{V}\rangle_{c6}|S_{H}\rangle_{c7}+|S_{H}\rangle_{c7}|L_{V}\rangle_{c8})+(1-t)|S_{H}L_{V}\rangle_{out3}\nonumber\\
&+&\frac{\sqrt{t(1-t)}}{\sqrt{2}}(|S_{H}\rangle_{c5}|L_{V}\rangle_{out3}-|S_{H}\rangle_{c7}|L_{V}\rangle_{out3}-|L_{V}\rangle_{c6}|S_{H}\rangle_{out3}-|L_{V}\rangle_{c8}|S_{H}\rangle_{out3})]\nonumber\\
&+&(\alpha|S_{H}\rangle_{b5}+\alpha|S_{H}\rangle_{b7}+\beta|L_{V}\rangle_{b6}+\beta|L_{V}\rangle_{b8})\nonumber\\
&\otimes&[\frac{t}{2}(|S_{H}\rangle_{a5}|L_{V}\rangle_{a6}-|S_{H}\rangle_{a5}|L_{V}\rangle_{a8}-|L_{V}\rangle_{a6}|S_{H}\rangle_{a7}+|S_{H}\rangle_{a7}|L_{V}\rangle_{a8})+(1-t)|S_{H}L_{V}\rangle_{out1}\nonumber\\
&+&\frac{\sqrt{t(1-t)}}{\sqrt{2}}(|S_{H}\rangle_{a5}|L_{V}\rangle_{out1}-|S_{H}\rangle_{a7}|L_{V}\rangle_{out1}-|L_{V}\rangle_{a6}|S_{H}\rangle_{out1}-|L_{V}\rangle_{a8}|S_{H}\rangle_{out1})]\nonumber\\
&\otimes&[\frac{t}{2}(|S_{H}\rangle_{b5}|L_{V}\rangle_{b6}-|S_{H}\rangle_{b5}|L_{V}\rangle_{b8}-|L_{V}\rangle_{b6}|S_{H}\rangle_{b7}+|S_{H}\rangle_{b7}|L_{V}\rangle_{b8})+(1-t)|S_{H}L_{V}\rangle_{out2}\nonumber\\
&+&\frac{\sqrt{t(1-t)}}{\sqrt{2}}(|S_{H}\rangle_{b5}|L_{V}\rangle_{out2}-|S_{H}\rangle_{b7}|L_{V}\rangle_{out2}-|L_{V}\rangle_{b6}|S_{H}\rangle_{out2}-|L_{V}\rangle_{b8}|S_{H}\rangle_{out2})]\nonumber\\
&\otimes&[\frac{t}{2}(|S_{H}\rangle_{c5}|L_{V}\rangle_{c6}-|S_{H}\rangle_{c5}|L_{V}\rangle_{c8}-|L_{V}\rangle_{c6}|S_{H}\rangle_{c7}+|S_{H}\rangle_{c7}|L_{V}\rangle_{c8})+(1-t)|S_{H}L_{V}\rangle_{out3}\nonumber\\
&+&\frac{\sqrt{t(1-t)}}{\sqrt{2}}(|S_{H}\rangle_{c5}|L_{V}\rangle_{out3}-|S_{H}\rangle_{c7}|L_{V}\rangle_{out3}-|L_{V}\rangle_{c6}|S_{H}\rangle_{out3}-|L_{V}\rangle_{c8}|S_{H}\rangle_{out3})]\nonumber\\
&+&(\alpha|S_{H}\rangle_{c5}+\alpha|S_{H}\rangle_{c7}+\beta|L_{V}\rangle_{c6}+\beta|L_{V}\rangle_{c8})\nonumber\\
&\otimes&[\frac{t}{2}(|S_{H}\rangle_{a5}|L_{V}\rangle_{a6}-|S_{H}\rangle_{a5}|L_{V}\rangle_{a8}-|L_{V}\rangle_{a6}|S_{H}\rangle_{a7}+|S_{H}\rangle_{a7}|L_{V}\rangle_{a8})+(1-t)|S_{H}L_{V}\rangle_{out1}\nonumber\\
&+&\frac{\sqrt{t(1-t)}}{\sqrt{2}}(|S_{H}\rangle_{a5}|L_{V}\rangle_{out1}-|S_{H}\rangle_{a7}|L_{V}\rangle_{out1}-|L_{V}\rangle_{a6}|S_{H}\rangle_{out1}-|L_{V}\rangle_{a8}|S_{H}\rangle_{out1})]\nonumber\\
&\otimes&[\frac{t}{2}(|S_{H}\rangle_{b5}|L_{V}\rangle_{b6}-|S_{H}\rangle_{b5}|L_{V}\rangle_{b8}-|L_{V}\rangle_{b6}|S_{H}\rangle_{b7}+|S_{H}\rangle_{b7}|L_{V}\rangle_{b8})+(1-t)|S_{H}L_{V}\rangle_{out2}\nonumber\\
&+&\frac{\sqrt{t(1-t)}}{\sqrt{2}}(|S_{H}\rangle_{b5}|L_{V}\rangle_{out2}-|S_{H}\rangle_{b7}|L_{V}\rangle_{out2}-|L_{V}\rangle_{b6}|S_{H}\rangle_{out2}-|L_{V}\rangle_{b8}|S_{H}\rangle_{out2})]\nonumber\\
&\otimes&[\frac{t}{2}(|S_{H}\rangle_{c5}|L_{V}\rangle_{c6}-|S_{H}\rangle_{c5}|L_{V}\rangle_{c8}-|L_{V}\rangle_{c6}|S_{H}\rangle_{c7}+|S_{H}\rangle_{c7}|L_{V}\rangle_{c8})+(1-t)|S_{H}L_{V}\rangle_{out3}\nonumber\\
&+&\frac{\sqrt{t(1-t)}}{\sqrt{2}}(|S_{H}\rangle_{c5}|L_{V}\rangle_{out3}-|S_{H}\rangle_{c7}|L_{V}\rangle_{out3}-|L_{V}\rangle_{c6}|S_{H}\rangle_{out3}-|L_{V}\rangle_{c8}|S_{H}\rangle_{out3})]\}.\label{whole3}
\end{eqnarray}

Then, the photons in all the twelve output modes are detected by the single-photon detectors, say, $D_{1a}$, $D_{2a}$, $D_{3a}$, $D_{4a}$, $D_{1b}$, $D_{2b}$, $D_{3b}$, $D_{4b}$, $D_{1c}$, $D_{2c}$, $D_{3c}$, and $D_{4c}$, respectively. According to the measurement results, we can judge whether the amplification protocol is successful. If each of the three parties obtains one of the following four measurement results, that is, $D_{1i}D_{2i}$, $D_{1i}D_{4i}$, $D_{2i}D_{3i}$, or $D_{3i}D_{4i}$ each registers one photon, where $i=a$, $b$, $c$, our protocol is successful. Otherwise, if the parties obtain other measurement results, the protocol fails. Therefore, there are totally sixty-four different measurement results corresponding to the successful cases.

    We take the measurement result of $D_{1a}D_{2a}D_{1b}D_{2b}D_{1c}D_{2c}$ for example. In Eq. (\ref{whole3}), there are six items which can make the detectors $D_{1a}D_{2a}D_{1b}D_{2b}D_{1c}D_{2c}$ each registers one photon.  They are $-\frac{\alpha t^{2}\sqrt{t(1-t)}}{8\sqrt{3}}|S_{H}\rangle_{a5}|L_{V}\rangle_{a6}|S_{H}\rangle_{b5}|L_{V}\rangle_{b6}|S_{H}\rangle_{c5}|L_{V}\rangle_{c6}|S_{H}\rangle_{out1}$,
   $\frac{\beta t^{2}\sqrt{t(1-t)}}{8\sqrt{3}}|S_{H}\rangle_{a5}|L_{V}\rangle_{a6}|S_{H}\rangle_{b5}|L_{V}\rangle_{b6}|S_{H}\rangle_{c5}|L_{V}\rangle_{c6}|L_{V}\rangle_{out1}$,  $-\frac{\alpha t^{2}\sqrt{t(1-t)}}{8\sqrt{3}}|S_{H}\rangle_{a5}|L_{V}\rangle_{a6}|S_{H}\rangle_{b5}|L_{V}\rangle_{b6}|S_{H}\rangle_{c5}|L_{V}\rangle_{c6}|S_{H}\rangle_{out2}$,  $\frac{\beta t^{2}\sqrt{t(1-t)}}{8\sqrt{3}}|S_{H}\rangle_{a5}|L_{V}\rangle_{a6}|S_{H}\rangle_{b5}|L_{V}\rangle_{b6}|S_{H}\rangle_{c5}|L_{V}\rangle_{c6}|L_{V}\rangle_{out2}$,  $-\frac{\alpha t^{2}\sqrt{t(1-t)}}{8\sqrt{3}}|S_{H}\rangle_{a5}|L_{V}\rangle_{a6}|S_{H}\rangle_{b5}|L_{V}\rangle_{b6}|S_{H}\rangle_{c5}|L_{V}\rangle_{c6}|S_{H}\rangle_{out3}$, and
    $\frac{\beta t^{2}\sqrt{t(1-t)}}{8\sqrt{3}}|S_{H}\rangle_{a5}|L_{V}\rangle_{a6}|S_{H}\rangle_{b5}|L_{V}\rangle_{b6}|S_{H}\rangle_{c5}|L_{V}\rangle_{c6}|L_{V}\rangle_{out3}$,   In this way, if the parties obtain the measurement result of $D_{1a}D_{2a}D_{1b}D_{2b}D_{1c}D_{2c}$, the state in Eq. (\ref{whole3}) will finally collapse to
\begin{eqnarray}
|\Phi_{1}\rangle_{ABC}&=&\frac{1}{\sqrt{3}}[(-\alpha|S_{H}\rangle_{out1}+\beta|L_{V}\rangle_{out1})|0\rangle_{out2}|0\rangle_{out3}+|0\rangle_{out1}(-\alpha|S_{H}\rangle_{out2}+\beta|L_{V}\rangle_{out2})|0\rangle_{out3}\nonumber\\
&+&|0\rangle_{out1}|0\rangle_{out2}(-\alpha|S_{H}\rangle_{out3}+\beta|L_{V}\rangle_{out3})],\label{new1}
\end{eqnarray}
with the probability of $\frac{t^{5}(1-t)}{64}$. $|\Phi_{1}\rangle_{ABC}$ can be easily transformed to $|\Phi\rangle_{ABC}$ in Eq. (\ref{max}) with the phase flip operation. Certainly, if the parties obtain the other sixty-three successful measurement results, they can also finally obtain the same state in Eq. (\ref{max}). Therefore, the total successful probability is $P_{1}=\frac{t^{5}(1-t)}{64}\times64=t^{5}(1-t)$.

On the other hand, if the initial photon state is the $|vac\rangle\otimes|\varphi_{ABC}\rangle$, when the parties make the photons in the $a_{1}a_{2}$, $b_{1}b_{2}$, and $c_{1}c_{2}$ modes pass through the BSs, they will obtain
\begin{eqnarray}
|\varphi_{ABC}\rangle&\rightarrow&[\frac{t}{2}(|S_{H}L_{V}\rangle_{a3}-|S_{H}L_{V}\rangle_{a3a4}-|L_{V}S_{H}\rangle_{a3a4}+|S_{H}L_{V}\rangle_{a4})+(1-t)|S_{H}L_{V}\rangle_{out1}\nonumber\\
&+&\frac{\sqrt{t(1-t)}}{\sqrt{2}}(|S_{H}L_{V}\rangle_{a3out1}-|S_{H}L_{V}\rangle_{a4out1}-|L_{V}S_{H}\rangle_{a3out1}-|L_{V}S_{H}\rangle_{a4out1})]\nonumber\\
&\otimes&[\frac{t}{2}(|S_{H}L_{V}\rangle_{b3}-|S_{H}L_{V}\rangle_{b3b4}-|L_{V}S_{H}\rangle_{b3b4}+|S_{H}L_{V}\rangle_{b4})
+(1-t)|S_{H}L_{V}\rangle_{out2}\nonumber\\
&+&\frac{\sqrt{t(1-t)}}{\sqrt{2}}(|S_{H}L_{V}\rangle_{b3out2}-|S_{H}L_{V}\rangle_{b4out2}-|L_{V}S_{H}\rangle_{b3out2}-|L_{V}S_{H}\rangle_{b4out2})]\nonumber\\
&\otimes&[\frac{t}{2}(|S_{H}L_{V}\rangle_{c3}-|S_{H}L_{V}\rangle_{c3c4}-|L_{V}S_{H}\rangle_{c3c4}+|S_{H}L_{V}\rangle_{c4})+(1-t)|S_{H}L_{V}\rangle_{out3}\nonumber\\
&+&\frac{\sqrt{t(1-t)}}{\sqrt{2}}(|S_{H}L_{V}\rangle_{c3out3}-|S_{H}L_{V}\rangle_{c4out3}-|L_{V}S_{H}\rangle_{c3out3}-|L_{V}S_{H}\rangle_{c4out3})].\label{whole4}
\end{eqnarray}

Next, the parties make the photons in the $a_{3}a_{4}$, $b_{3}b_{4}$, and $c_{3}c_{4}$ modes pass through $PBS_{3a}PBS_{4a}$, $PBS_{3b}PBS_{4b}$, and $PBS_{3c}PBS_{4c}$, respectively. After the PBSs, the state in Eq. (\ref{whole4}) will evolve to
\begin{eqnarray}
|\varphi_{ABC}\rangle&\rightarrow&[\frac{t}{2}(|S_{H}\rangle_{a5}|L_{V}\rangle_{a6}-|S_{H}\rangle_{a5}|L_{V}\rangle_{a8}-|L_{V}\rangle_{a6}|S_{H}\rangle_{a7}+|S_{H}\rangle_{a7}|L_{V}\rangle_{a8})+(1-t)|S_{H}L_{V}\rangle_{out1}\nonumber\\
&+&\frac{\sqrt{t(1-t)}}{\sqrt{2}}(|S_{H}\rangle_{a5}|L_{V}\rangle_{out1}-|S_{H}\rangle_{a7}|L_{V}\rangle_{out1}-|L_{V}\rangle_{a6}|S_{H}\rangle_{out1}-|L_{V}\rangle_{a8}|S_{H}\rangle_{out1})]\nonumber\\
&\otimes&[\frac{t}{2}(|S_{H}\rangle_{b5}|L_{V}\rangle_{b6}-|S_{H}\rangle_{b5}|L_{V}\rangle_{b8}-|L_{V}\rangle_{b6}|S_{H}\rangle_{b7}+|S_{H}\rangle_{b7}|L_{V}\rangle_{b8})+(1-t)|S_{H}L_{V}\rangle_{out2}\nonumber\\
&+&\frac{\sqrt{t(1-t)}}{\sqrt{2}}(|S_{H}\rangle_{b5}|L_{V}\rangle_{out2}-|S_{H}\rangle_{b7}|L_{V}\rangle_{out2}-|L_{V}\rangle_{b6}|S_{H}\rangle_{out2}-|L_{V}\rangle_{b8}|S_{H}\rangle_{out2})]\nonumber\\
&\otimes&[\frac{t}{2}(|S_{H}\rangle_{c5}|L_{V}\rangle_{c6}-|S_{H}\rangle_{c5}|L_{V}\rangle_{c8}-|L_{V}\rangle_{c6}|S_{H}\rangle_{c7}+|S_{H}\rangle_{c7}|L_{V}\rangle_{c8})+(1-t)|S_{H}L_{V}\rangle_{out3}\nonumber\\
&+&\frac{\sqrt{t(1-t)}}{\sqrt{2}}(|S_{H}\rangle_{c5}|L_{V}\rangle_{out3}-|S_{H}\rangle_{c7}|L_{V}\rangle_{out3}-|L_{V}\rangle_{c6}|S_{H}\rangle_{out3}-|L_{V}\rangle_{c8}|S_{H}\rangle_{out3})].\label{whole5}
\end{eqnarray}
 We can prove that if the parties obtain one of the above successful measurement results, the state in Eq. (\ref{whole5}) will collapse to the vacuum state. For example, in Eq. (\ref{whole5}), only the item $\frac{t^{3}}{8}|S_{H}\rangle_{a5}|L_{V}\rangle_{a6}|S_{H}\rangle_{b5}|L_{V}\rangle_{b6}|S_{H}\rangle_{c5}|L_{V}\rangle_{c6}$ can make the detectors $D_{1a}D_{2a}D_{1b}D_{2b}D_{1c}D_{2c}$ each registers one photon. In this way, under the measurement result of $D_{1a}D_{2a}D_{1b}D_{2b}D_{1c}D_{2c}$, the state in Eq. (\ref{whole5}) will collapse to the vacuum state with the probability of $\frac{t^{6}}{64}$. Similarly, if the parties obtain the other sixty-three successful measurement results, they can also finally obtain the vacuum state. Therefore, the total successful probability to obtain the vacuum state is $P_{2}=\frac{t^{6}}{64}\times64=t^{6}$.

According to the description above, if the protocol is successful, the parties can obtain a new mixed state as
\begin{eqnarray}
\rho_{out}=\eta'|\Phi\rangle_{ABC}\langle\Phi|+(1-\eta')|vac\rangle\langle vac|,
\end{eqnarray}
where the fidelity $\eta'$ is
\begin{eqnarray}
\eta'=\frac{\eta P_{1}}{\eta P_{1}+(1-\eta)P_{2}}=\frac{\eta(1-t)}{\eta-2\eta t+t}.\label{fidelity}
\end{eqnarray}
The total successful probability for our protocol is
\begin{eqnarray}
P_{t}=\eta P_{1}+(1-\eta)P_{2}=t^{5}(\eta-2\eta t+t).
\end{eqnarray}
It can be found $\rho_{out}$ has the same form of $\rho_{in}$. The entanglement coefficients $\alpha$ and $\beta$ do not effect the $P_{t}$ and $\eta'$, which are only functions of the initial fidelity $\eta$ and the transmission $t$. We define the amplification factor $g$ as
\begin{eqnarray}
g\equiv\frac{\eta'}{\eta}=\frac{1-t}{\eta (1-t)+(1-\eta)t}. \label{g}
\end{eqnarray}
For realizing the amplification, we require $g>1$. It can be calculated that $g>1$ under the case of $t<\frac{1}{2}$. In this way, by providing suitable VBSs with $t<\frac{1}{2}$, we can complete the amplification task.

\section{The heralded amplification for the single-photon N-mode W state of the time-bin qubit}
\begin{figure}[!h]
\begin{center}
\includegraphics[width=16cm,angle=0]{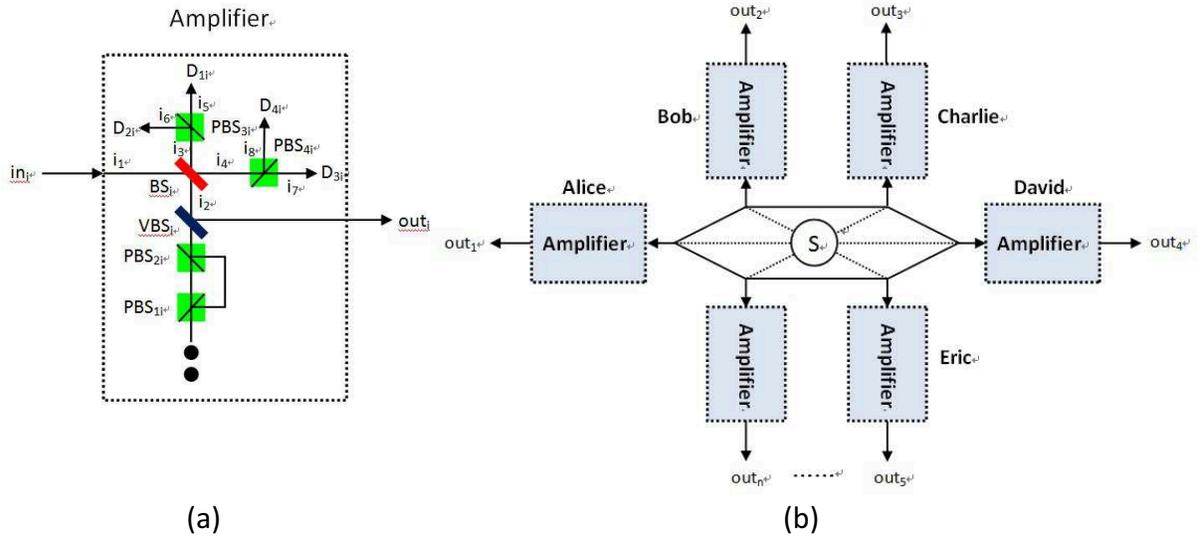}
\caption{(a) The structure of the "Amplifier". (b) The schematic principle of the amplification protocol for the single-photon N-mode W state of the time-bin qubit. }
\end{center}
\end{figure}

Our protocol can be straightly extended to single-photon N-mode W state of the time-bin qubit, with $N>3$. Suppose a time-bin qubit with the form of $|\psi\rangle$ in Eq. (\ref{timebin}) is distributed to N parties, say, Alice (A), Bob (B), Charlie (C), David (D), $\cdots$, which creates a single-photon N-mode W state as
\begin{eqnarray}
|\Phi\rangle_{N}=\frac{1}{\sqrt{N}}(|\psi\rangle_{a1}|0\rangle_{b1}|0\rangle_{c1}\cdots|0\rangle_{n1}+|0\rangle_{a1}|\psi\rangle_{b1}|0\rangle_{c1}\cdots|0\rangle_{n1}
+\cdots+|0\rangle_{a1}|0\rangle_{b1}\cdots|\psi\rangle_{n1}).\label{maxN}
\end{eqnarray}
We consider the single-qubit can be completely lost with the probability of $1-\eta_{N}$. In this way, the three parties share a mixed state as
 \begin{eqnarray}
\rho_{N}=\eta_{N}|\Phi\rangle_{N}\langle\Phi|+(1-\eta_{N})|vac\rangle\langle vac|.
\end{eqnarray}

The schematic principle of our protocol is shown in Fig. 2 (b), and the structure of "Amplifier" is shown in Fig. 2 (a), where $i=a$, $b$, $c$, $d$, $\cdots$, $n$. Each of the $N$ parties needs to prepare two auxiliary photons with the polarization of $|H\rangle$ and $|V\rangle$, respectively. They each make the two photons pass through two PBSs and create the auxiliary photon state as $|\varphi\rangle=|S_{H}\rangle\otimes|L_{V}\rangle$.
Then,  each of them makes the auxiliary photons in his or her hand pass through one VBS with the transmission of $t$. After the VBSs, they can obtain
\begin{eqnarray}
|\varphi_{N}\rangle&=&[t|S_{H}L_{V}\rangle_{a2}+(1-t)|S_{H}L_{V}\rangle_{out1}+\sqrt{t(1-t)}(|S_{H}L_{V}\rangle_{a2out1}+|L_{V}S_{H}\rangle_{a2out1})]\nonumber\\
&\otimes&[t|S_{H}L_{V}\rangle_{b2}+(1-t)|S_{H}L_{V}\rangle_{out2}+\sqrt{t(1-t)}(|S_{H}L_{V}\rangle_{b2out2}+|L_{V}S_{H}\rangle_{b2out2})]\nonumber\\
&\otimes&[t|S_{H}L_{V}\rangle_{c2}+(1-t)|S_{H}L_{V}\rangle_{out3}+\sqrt{t(1-t)}(|S_{H}L_{V}\rangle_{c2out3}+|L_{V}S_{H}\rangle_{c2out3})]\nonumber\\
&&\cdots\nonumber\\
&\otimes&[t|S_{H}L_{V}\rangle_{n2}+(1-t)|S_{H}L_{V}\rangle_{outn}+\sqrt{t(1-t)}(|S_{H}L_{V}\rangle_{n2outn}+|L_{V}S_{H}\rangle_{n2outn})].
\end{eqnarray}

Each of the N parties makes the photons in the $i_{1}i_{2}$ modes pass through the BS, and then make all the output photons pass through the PBSs.  If the whole photon state is $|\Phi\rangle_{N}\otimes|\varphi_{N}\rangle$ with the probability of $\eta_{N}$, it will evolve to
\begin{eqnarray}
|\Phi\rangle_{N}\otimes|\varphi_{N}\rangle&\rightarrow&\frac{1}{\sqrt{2N}}\{(\alpha|S_{H}\rangle_{a5}+\alpha|S_{H}\rangle_{a7}+\beta|L_{V}\rangle_{a6}+\beta|L_{V}\rangle_{a8})\nonumber\\
&\otimes&[\frac{t}{2}(|S_{H}\rangle_{a5}|L_{V}\rangle_{a6}-|S_{H}\rangle_{a5}|L_{V}\rangle_{a8}-|L_{V}\rangle_{a6}|S_{H}\rangle_{a7}+|S_{H}\rangle_{a7}|L_{V}\rangle_{a8})+(1-t)|S_{H}L_{V}\rangle_{out1}\nonumber\\
&+&\frac{\sqrt{t(1-t)}}{\sqrt{2}}(|S_{H}\rangle_{a5}|L_{V}\rangle_{out1}-|S_{H}\rangle_{a7}|L_{V}\rangle_{out1}-|L_{V}\rangle_{a6}|S_{H}\rangle_{out1}-|L_{V}\rangle_{a8}|S_{H}\rangle_{out1})]\nonumber\\
&\otimes&[\frac{t}{2}(|S_{H}\rangle_{b5}|L_{V}\rangle_{b6}-|S_{H}\rangle_{b5}|L_{V}\rangle_{b8}-|L_{V}\rangle_{b6}|S_{H}\rangle_{b7}+|S_{H}\rangle_{b7}|L_{V}\rangle_{b8})+(1-t)|S_{H}L_{V}\rangle_{out2}\nonumber\\
&+&\frac{\sqrt{t(1-t)}}{\sqrt{2}}(|S_{H}\rangle_{b5}|L_{V}\rangle_{out2}-|S_{H}\rangle_{b7}|L_{V}\rangle_{out2}-|L_{V}\rangle_{b6}|S_{H}\rangle_{out2}-|L_{V}\rangle_{b8}|S_{H}\rangle_{out2})]\nonumber\\
&\otimes&\cdots\nonumber\\
&\otimes&[\frac{t}{2}(|S_{H}\rangle_{n5}|L_{V}\rangle_{n6}-|S_{H}\rangle_{n5}|L_{V}\rangle_{n8}-|L_{V}\rangle_{n6}|S_{H}\rangle_{n7}+|S_{H}\rangle_{n7}|L_{V}\rangle_{n8})+(1-t)|S_{H}L_{V}\rangle_{outn}\nonumber\\
&+&\frac{\sqrt{t(1-t)}}{\sqrt{2}}(|S_{H}\rangle_{n5}|L_{V}\rangle_{outn}-|S_{H}\rangle_{n7}|L_{V}\rangle_{outn}-|L_{V}\rangle_{n6}|S_{H}\rangle_{outn}-|L_{V}\rangle_{n8}|S_{H}\rangle_{outn})]\nonumber\\
&+&(\alpha|S_{H}\rangle_{b5}+\alpha|S_{H}\rangle_{b7}+\beta|L_{V}\rangle_{b6}+\beta|L_{V}\rangle_{b8})\nonumber\\
&\otimes&[\frac{t}{2}(|S_{H}\rangle_{a5}|L_{V}\rangle_{a6}-|S_{H}\rangle_{a5}|L_{V}\rangle_{a8}-|L_{V}\rangle_{a6}|S_{H}\rangle_{a7}+|S_{H}\rangle_{a7}|L_{V}\rangle_{a8})+(1-t)|S_{H}L_{V}\rangle_{out1}\nonumber\\
&+&\frac{\sqrt{t(1-t)}}{\sqrt{2}}(|S_{H}\rangle_{a5}|L_{V}\rangle_{out1}-|S_{H}\rangle_{a7}|L_{V}\rangle_{out1}-|L_{V}\rangle_{a6}|S_{H}\rangle_{out1}-|L_{V}\rangle_{a8}|S_{H}\rangle_{out1})]\nonumber\\
&\otimes&[\frac{t}{2}(|S_{H}\rangle_{b5}|L_{V}\rangle_{b6}-|S_{H}\rangle_{b5}|L_{V}\rangle_{b8}-|L_{V}\rangle_{b6}|S_{H}\rangle_{b7}+|S_{H}\rangle_{b7}|L_{V}\rangle_{b8})+(1-t)|S_{H}L_{V}\rangle_{out2}\nonumber\\
&+&\frac{\sqrt{t(1-t)}}{\sqrt{2}}(|S_{H}\rangle_{b5}|L_{V}\rangle_{out2}-|S_{H}\rangle_{b7}|L_{V}\rangle_{out2}-|L_{V}\rangle_{b6}|S_{H}\rangle_{out2}-|L_{V}\rangle_{b8}|S_{H}\rangle_{out2})]\nonumber\\
&\otimes&\cdots\nonumber\\
&\otimes&[\frac{t}{2}(|S_{H}\rangle_{n5}|L_{V}\rangle_{n6}-|S_{H}\rangle_{n5}|L_{V}\rangle_{n8}-|L_{V}\rangle_{n6}|S_{H}\rangle_{n7}+|S_{H}\rangle_{n7}|L_{V}\rangle_{n8})+(1-t)|S_{H}L_{V}\rangle_{outn}\nonumber\\
&+&\frac{\sqrt{t(1-t)}}{\sqrt{2}}(|S_{H}\rangle_{n5}|L_{V}\rangle_{outn}-|S_{H}\rangle_{n7}|L_{V}\rangle_{outn}-|L_{V}\rangle_{n6}|S_{H}\rangle_{outn}-|L_{V}\rangle_{n8}|S_{H}\rangle_{outn})]\nonumber\\
&+&\cdots\nonumber\\
&+&(\alpha|S_{H}\rangle_{n5}+\alpha|S_{H}\rangle_{n7}+\beta|L_{V}\rangle_{n6}+\beta|L_{V}\rangle_{n8})\nonumber\\
&\otimes&[\frac{t}{2}(|S_{H}\rangle_{a5}|L_{V}\rangle_{a6}-|S_{H}\rangle_{a5}|L_{V}\rangle_{a8}-|L_{V}\rangle_{a6}|S_{H}\rangle_{a7}+|S_{H}\rangle_{a7}|L_{V}\rangle_{a8})+(1-t)|S_{H}L_{V}\rangle_{out1}\nonumber\\
&+&\frac{\sqrt{t(1-t)}}{\sqrt{2}}(|S_{H}\rangle_{a5}|L_{V}\rangle_{out1}-|S_{H}\rangle_{a7}|L_{V}\rangle_{out1}-|L_{V}\rangle_{a6}|S_{H}\rangle_{out1}-|L_{V}\rangle_{a8}|S_{H}\rangle_{out1})]\nonumber\\
&\otimes&[\frac{t}{2}(|S_{H}\rangle_{b5}|L_{V}\rangle_{b6}-|S_{H}\rangle_{b5}|L_{V}\rangle_{b8}-|L_{V}\rangle_{b6}|S_{H}\rangle_{b7}+|S_{H}\rangle_{b7}|L_{V}\rangle_{b8})+(1-t)|S_{H}L_{V}\rangle_{out2}\nonumber\\
&+&\frac{\sqrt{t(1-t)}}{\sqrt{2}}(|S_{H}\rangle_{b5}|L_{V}\rangle_{out2}-|S_{H}\rangle_{b7}|L_{V}\rangle_{out2}-|L_{V}\rangle_{b6}|S_{H}\rangle_{out2}-|L_{V}\rangle_{b8}|S_{H}\rangle_{out2})]\nonumber\\
&\otimes&\cdots\nonumber\\
&\otimes&[\frac{t}{2}(|S_{H}\rangle_{n5}|L_{V}\rangle_{n6}-|S_{H}\rangle_{n5}|L_{V}\rangle_{n8}-|L_{V}\rangle_{n6}|S_{H}\rangle_{n7}+|S_{H}\rangle_{n7}|L_{V}\rangle_{n8})+(1-t)|S_{H}L_{V}\rangle_{outn}\nonumber\\
&+&\frac{\sqrt{t(1-t)}}{\sqrt{2}}(|S_{H}\rangle_{n5}|L_{V}\rangle_{outn}-|S_{H}\rangle_{n7}|L_{V}\rangle_{outn}-|L_{V}\rangle_{n6}|S_{H}\rangle_{outn}-|L_{V}\rangle_{n8}|S_{H}\rangle_{outn})]\}.\label{wholeN2}
\end{eqnarray}

Finally, the parties measure the photons in all the $4N$ output modes by the single-photon detectors. If each of the $N$ parties obtains one of the following four measurement results, say, $D_{1i}D_{2i}$, $D_{1i}D_{4i}$, $D_{2i}D_{3i}$, or $D_{3i}D_{4i}$ each registers one photon, where $i=a$, $b$, $c$, $\cdots$, $n$, our protocol is successful. Otherwise, if the parties obtain other measurement results, the protocol fails. Therefore, there are totally $4^{N}$ different successful measurement results. For example, if each of the parties obtain the measurement result of $D_{1i}D_{2i}$ each registers one photon, where $i=a$, $b$, $c$, $\cdots$, $n$, the state in Eq. (\ref{wholeN2}) will collapse to
\begin{eqnarray}
|\Phi_{N1}\rangle&=&\frac{1}{\sqrt{N}}[(-\alpha|S_{H}\rangle_{out1}+\beta|L_{V}\rangle_{out1})|0\rangle_{out2}|0\rangle_{out3}\cdots|0\rangle_{outn}+|0\rangle_{out1}(-\alpha|S_{H}\rangle_{out2}+\beta|L_{V}\rangle_{out2})|0\rangle_{out3}\cdots|0\rangle_{outn}\nonumber\\
&+&\cdots+|0\rangle_{out1}|0\rangle_{out2}\cdots(-\alpha|S_{H}\rangle_{outn}+\beta|L_{V}\rangle_{outn})],\label{new1}
\end{eqnarray}
with the probability of $\frac{t^{2N-1}(1-t)}{4^{N}}$. $|\Phi_{N1}\rangle$ can be converted to $|\Phi\rangle_{N}$ in Eq. (\ref{maxN}) with the phase-flip operation. If the parties obtain other successful measurement results, they can also obtain the same state in Eq. (\ref{maxN}) with the help of the phase-flip operation. In this way, the total successful probability is $P_{1N}=\frac{t^{2N-1}(1-t)}{4^{N}}\times4^{N}=t^{2N-1}(1-t)$.

  Under the condition that the initial photon state is the vacuum state, when the parties make the photons in the $i_{1}i_{2}$ ($i=a$, $b$, $c$, $\cdots$, $n$ )modes pass through the BSs and the output photons pass through the PBS, they will obtain
\begin{eqnarray}
|\varphi_{N}\rangle&\rightarrow&[\frac{t}{2}(|S_{H}\rangle_{a5}|L_{V}\rangle_{a6}-|S_{H}\rangle_{a5}|L_{V}\rangle_{a8}-|L_{V}\rangle_{a6}|S_{H}\rangle_{a7}+|S_{H}\rangle_{a7}|L_{V}\rangle_{a8})+(1-t)|S_{H}L_{V}\rangle_{out1}\nonumber\\
&+&\frac{\sqrt{t(1-t)}}{\sqrt{2}}(|S_{H}\rangle_{a5}|L_{V}\rangle_{out1}-|S_{H}\rangle_{a7}|L_{V}\rangle_{out1}-|L_{V}\rangle_{a6}|S_{H}\rangle_{out1}-|L_{V}\rangle_{a8}|S_{H}\rangle_{out1})]\nonumber\\
&\otimes&[\frac{t}{2}(|S_{H}\rangle_{b5}|L_{V}\rangle_{b6}-|S_{H}\rangle_{b5}|L_{V}\rangle_{b8}-|L_{V}\rangle_{b6}|S_{H}\rangle_{b7}+|S_{H}\rangle_{b7}|L_{V}\rangle_{b8})+(1-t)|S_{H}L_{V}\rangle_{out2}\nonumber\\
&+&\frac{\sqrt{t(1-t)}}{\sqrt{2}}(|S_{H}\rangle_{b5}|L_{V}\rangle_{out2}-|S_{H}\rangle_{b7}|L_{V}\rangle_{out2}-|L_{V}\rangle_{b6}|S_{H}\rangle_{out2}-|L_{V}\rangle_{b8}|S_{H}\rangle_{out2})]\nonumber\\
&\otimes&\cdots\nonumber\\
&\otimes&[\frac{t}{2}(|S_{H}\rangle_{n5}|L_{V}\rangle_{n6}-|S_{H}\rangle_{n5}|L_{V}\rangle_{n8}-|L_{V}\rangle_{n6}|S_{H}\rangle_{n7}+|S_{H}\rangle_{n7}|L_{V}\rangle_{n8})+(1-t)|S_{H}L_{V}\rangle_{outn}\nonumber\\
&+&\frac{\sqrt{t(1-t)}}{\sqrt{2}}(|S_{H}\rangle_{n5}|L_{V}\rangle_{outn}-|S_{H}\rangle_{n7}|L_{V}\rangle_{outn}-|L_{V}\rangle_{n6}|S_{H}\rangle_{outn}-|L_{V}\rangle_{n8}|S_{H}\rangle_{outn})].\label{wholeN3}
\end{eqnarray}
 If the parties obtain one of the above successful measurement results, the state in Eq. (\ref{wholeN3}) will finally collapse to the vacuum state. The total successful probability to obtain the vacuum state is $P_{2N}=\frac{t^{2N}}{4^{N}}\times 4^{N}=t^{2N}$.

Therefore, for the the single-photon N-mode W state, if the protocol is successful, the parties can obtain a new mixed state as
\begin{eqnarray}
\rho_{outN}=\eta_{N}'|\Phi\rangle_{N}\langle\Phi|+(1-\eta_{N}')|vac\rangle\langle vac|,
\end{eqnarray}
where the fidelity $\eta_{N}'$ is
\begin{eqnarray}
\eta_{N}'=\frac{\eta_{N} P_{1N}}{\eta P_{1N}+(1-\eta_{N})P_{2N}}=\frac{\eta_{N}(1-t)}{\eta_{N}-2\eta_{N} t+t}.\label{fidelityn}
\end{eqnarray}
In this way, the amplification factor $g_{N}$ can be written as
\begin{eqnarray}
g_{N}=\frac{\eta_{N}'}{\eta_{N}}=\frac{1-t}{\eta_{N} (1-t)+(1-\eta_{N})t},\label{gn}
\end{eqnarray}
which has the same form of $g$ in Eq. (\ref{g}). It indicates that the mode number $N$ would not affect the fidelity of the distilled new mixed state. By providing the VBSs with $t<\frac{1}{2}$, we can obtain $g_{N}>1$ and realize the amplification.

The total successful probability for our protocol is
\begin{eqnarray}
P_{tN}=\eta_{N} P_{1N}+(1-\eta_{N})P_{2N}=t^{2N-1}(\eta_{N}-2\eta_{N} t+t).\label{pn}
\end{eqnarray}
It can be found that the value of $N$ would reduce the success probability of our protocol. As $t<1$, the larger value of $N$ leads to the lower $P_{tN}$.

\section{Disccusion and conclusion}
\begin{figure}[!h]
\begin{center}
\includegraphics[width=7cm,angle=0]{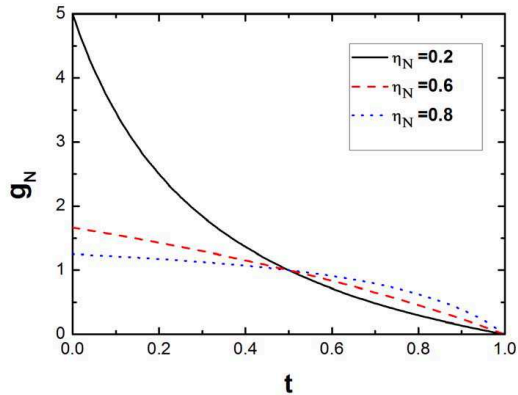}
\caption{The amplification factor $g_{N}$ as a function of the transmission $t$ of the VBSs under the initial fidelity $\eta=0.2$, $0.6$, and $0.8$, respectively. All three curves pass through the same point with $t=0.5$. }
\end{center}
\end{figure}

In this paper, we put forward an effective amplification protocol for the arbitrary single-photon N-mode W state of the time-bin qubit. We suppose one time-bin qubit is shared by N parties, which creates a single-photon N-mode W state. Due to the photon loss, the entangled state would be mixed with the vacuum state. For increasing the fidelity of the single-photon N-mode W state, each of the N parties requires to prepare two single photons with the polarization of $|H\rangle$ and $|V\rangle$, respectively. With the help of the PBSs, the polarization modes can accompany two temporal modes, respectively. Then, with the help of the local operation and some linear optical elements, such as the BSs, PBSs, and VBSs with the transmission of $t$, the parties can distill a new mixed state with the similar form of the initial mixed state with some probability. Under the case that $t<\frac{1}{2}$, the amplification factor $g\equiv\frac{\eta_{N}'}{\eta_{N}}>1$, which indicates the fidelity of the new mixed state is higher than that of the initial mixed state. Our protocol has three obvious advantages.
First, the parties only require one pair of single-photon N-mode W state of the time-bin qubit. As the entanglement source
is quite precious, our protocol is quite economical. Second, the information encoded in the time-bin qubit can be well remained after the amplification. Third, the protocol only requires the linear optical
elements, which makes it can be realized in current experimental
conditions.

Next, we discuss the experimental realization of our protocol. Our protocol relies on three kinds of beam splitters, say VBSs, PBSs and BSs, which are common linear optical elements in current technology. The VBS is the key element of our protocol. In our protocol, for realizing the amplification, we require to use the VBSs with $t<\frac{1}{2}$. The experimental transmission adjustion of the VBS has been reported by Osorio \emph{et al.} in 2012 \cite{NLA6}. They reported their experimental results
about the heralded photon amplification for quantum communication
with the help of the VBS . In the experiment, they successfully adjusted the splitting
ratio of VBS from $50:50$ to $90:10$ to increase the visibility from
$46.7\pm 3.1\%$ to $96.3\pm 3.8\%$. Based on their experimental result, our requirement for the VBS can be easily realized. Meanwhile, we also require the sophisticated single photon detectors to exactly distinguish
the single photon in each output modes. The single photon detection has been a challenge under current experimental
conditions, for the quantum decoherence effect of the photon detector \cite{photonefficiency}. Fortunately, the group of Lita once reported their experimental
result about the near-infrared single-photon detection. They showed the photon detection efficiency $\eta_{p}$ at 1556 $nm$ can reach $95\% \pm 2\%$ \cite{photonefficiency1}.

Finally, we numerically calculate the amplification factor $g_{N}$ and the success probability $P_{tN}$ of our protocol. As shown in Eq. (\ref{gn}), the amplification factor $g_{N}$ relies on the value of $t$ and the fidelity ($\eta_{N}$) of the initial mixed state. We show the value of $g_{N}$ as a function of  $t$ under different values of $\eta_{N}$ in Fig. 3. It can be found that $g_{N}$ reduces with the growth of $t$. All the curves  pass through the same point of $t=\frac{1}{2}$. Under $t=\frac{1}{2}$, we can obtain $g_{N}=1$ under any initial fidelity $\eta_{N}$. Actually, under $t=\frac{1}{2}$, the VBS becomes the BS and our protocol is analogous to an entanglement swapping protocol. Under $t<\frac{1}{2}$, we can obtain $g_{N}>1$. Meanwhile, under $t\rightarrow 0$, $g_{N}\rightarrow\frac{1}{\eta}$, which indicates that we can obtain $\eta'_{N}\rightarrow 1$ under the condition of  $t\rightarrow 0$. In this way, for obtaining high fidelity, the parties require to choose the VBS with small transmission.

\begin{figure}[!h]
\begin{center}
\includegraphics[width=7cm,angle=0]{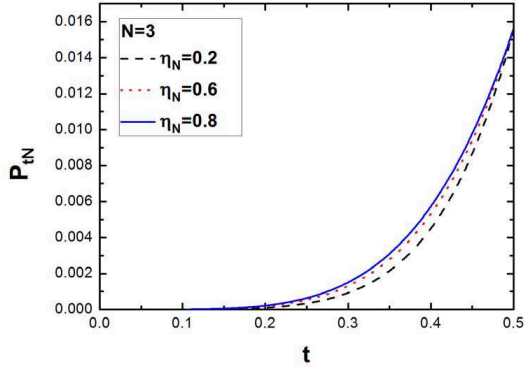}
\caption{The total success probability $P_{t}$ of our amplification protocol as a function of the transmission $t$ of the VBSs under the mode number $N=3$ and the initial fidelity $\eta=0.2$, $0.6$, and $0.8$, respectively. }
\end{center}
\end{figure}

\begin{figure}[!h]
\begin{center}
\includegraphics[width=7cm,angle=0]{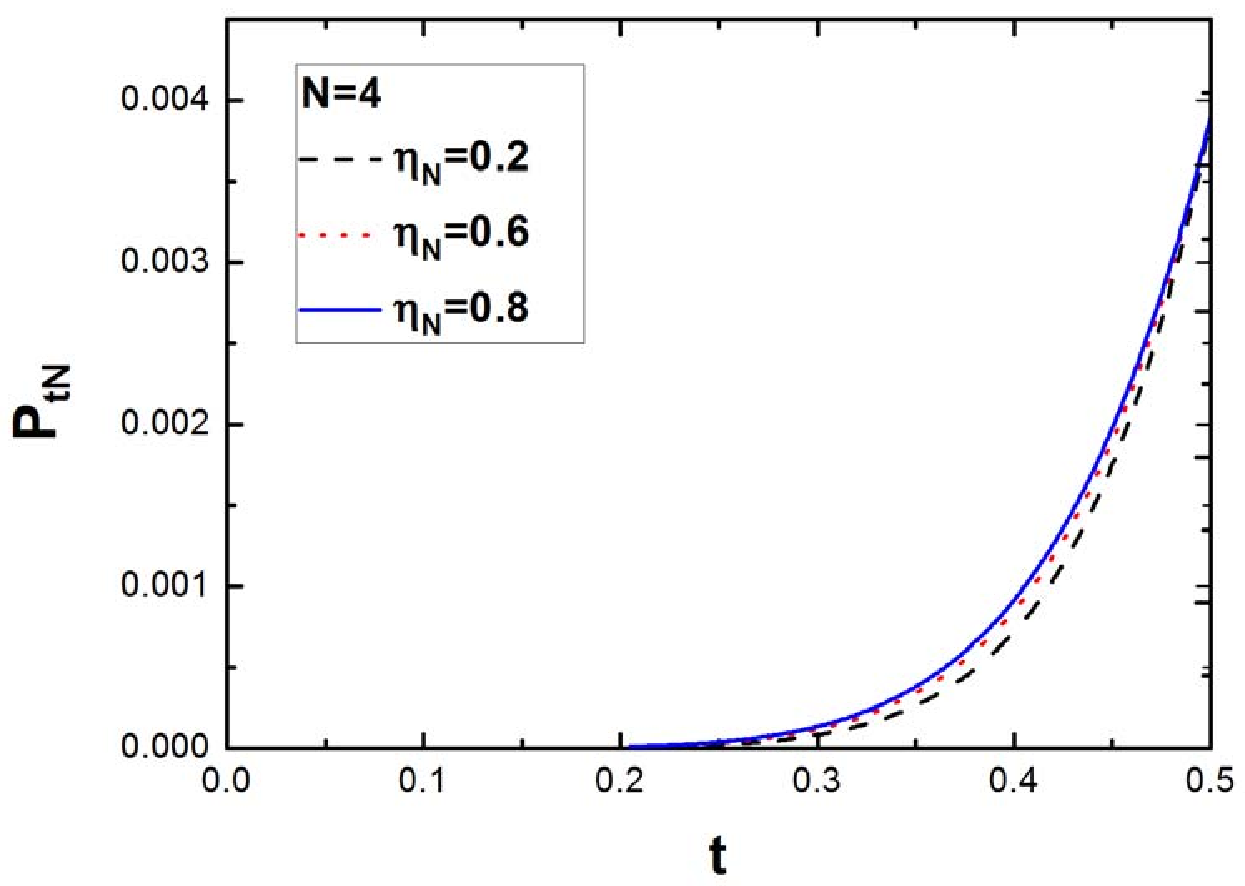}
\caption{The total success probability $P_{t}$ of our amplification protocol as a function of the transmission $t$ of the VBSs under the mode number $N=4$ anad the initial fidelity $\eta=0.2$, $0.6$, and $0.8$, respectively. }
\end{center}
\end{figure}

On the other hand, as shown in Eq. (\ref{pn}), the total success probability of our protocol relies on $t$, $\eta_{N}$ and the mode number $N$. Fig. 4 and Fig. 5 show the value of $P_{tN}$ as a function of $t$ with three different initial $\eta=0.2$, $0.6$, and $0.8$, under the condition of $N=3$ and $N=4$, respectively. It can be found that  the value of $P_{tN}$ mainly depends on $t$ and $N$, and the value of $\eta_{N}$ affects $P_{tN}$ slightly. $P_{tN}$ increases with the growth of $t$ while decreaases with the growth of $N$. Under $t\in [0,\frac{1}{2}]$, $P_{tN}$ will get the maximum value of $t^{2N}$ when $t=\frac{1}{2}$. When $t\rightarrow 0$, $P_{tN}\rightarrow 0$. In this way, in the practical application, we need to consider both the amplification factor and success probability factor, simultaneously, and choose the VBSs with suitable transmission.

In conclusion, we demonstrate an effective amplification protocol for protecting the single-photon multi-mode W state of the time-bin qubit. In the protocol, only one pair of the single-photon multi-mode W state and some auxiliary single photons are required, which makes our protocol economical. With the help of the auxiliary single photons and the linear optical elements, such as VBSs, BSs, and PBSs, we can successfully distill new mixed state, and the fidelity of the new mixed state can be effectively increased when the transmission $t$ of the VBSs satisfy $t<\frac{1}{2}$. Moreover, the encoded time-bin feature can be well contained. This protocol can be realized under current experimental condition, and it may be useful in current and future quantum information processing.

\section*{ACKNOWLEDGEMENTS} This work was supported by the National Natural Science Foundation
of China under Grant  Nos. 11474168 and 61401222, the Natural Science Foundation of Jiangsu province under Grant No. BK20151502
, the Qing Lan Project in Jiangsu Province, and A Project
Funded by the Priority Academic Program Development of Jiangsu
Higher Education Institutions.


\begin{thebibliography}{99}

\bibitem{teleportation} C. H. Bennett, G. Brassard, C. Crepeau,  R. Jozsa,
 and A. Peres, W. K. Wootters,  "Teleporting an unknown quantum state via dual classical and
Einstein-Podolsky-Rosen channels," Phys. Rev. Lett. \textbf{70}, 1895 (1993)

\bibitem{teleportation1} A. N. Pyrkov and T. Byrnes, "Quantum teleportation of spin coherent states: beyond continuous variables teleportation,"  New J. Phys. \textbf{16}, 073038 (2014)

\bibitem{Ekert91}  A. K. Ekert, "Quantum cryptography based on Bells theorem," Phys. Rev. Lett. \textbf{67}, 661 (1991)

\bibitem{qkd1} G. Vallone, A. Dall'Arche, M. Tomasin, and P. Villoresi, "Loss tolerant device-independent quantum key distribution: a proof of principle," New J. Phys. \textbf{16}, 063064 (2014)

\bibitem{qkd2} D. S. Simon and A. V. Sergienko, "High-capacity quantum key distribution via hyperentangled degrees of freedom," New J. Phys. \textbf{16}, 063052 (2014)


\bibitem{QSS} M. Hillery, V. Bu$\breve{z}$ek, and A. Berthiaume, "Quantum secret sharing," Phys. Rev. A
\textbf{59}, 1829 (1999)

\bibitem{QSDC1}G. L. Long, and X. S. Liu, "Theoretically efficient high-capacity quantum-keydistribution
scheme," Phys. Rev. A \textbf{65}, 032302 (2002)


\bibitem{QSDC2} F. G. Deng, G. L. Long, and X. S. Liu, "Two-step quantum direct communication
protocol using the Einstein-Podolsky-Rosen pair block," Phys. Rev. A \textbf{68},
042317 (2003)

\bibitem{dengcomputation4} H. R. Wei, and F. G. Deng, "Universal quantum gates on electron-spin qubits with quantum dots inside single-side optical microcavities". Opt. Exp. \textbf{22}, 593-607 (2014)

\bibitem{dengcomputation5}H. R. Wei, and F. G. Deng, "Scalable photonic quantum computing assisted by quantum-dot spin in double-sided optical microcavity," Opt. Exp. \textbf{21}, 17671-17685 (2013)

\bibitem{wangcomputation}C. Wang, Y. Zhang, R. Z. Jiao, G. S. Jin, "Universal quantum controlled phase gate on photonic qubits based on nitrogen vacancy centers and microcavity resonators," Opt. Exp. \textbf{21}, 19525-19260 (2013)

\bibitem{cryptography1} C. Silberhorn, T. C. Ralph, N. L$\ddot{u}$tkenhaus, and G. Leuchs,
"Continuous variable quantum cryptography: beating the 3 dB
loss limit," Phys. Rev. Lett. 89, 167901每167904 (2002).

\bibitem{cryptography2} Ch. Silberhorn, N. Korolkova, and G. Leuchs, "Quantum key
distribution with bright entangled beams," Phys. Rev. Lett. 88, 167902每167905 (2002).

\bibitem{engineering} M. G. A. Paris, M. Cola, and R. Bonifacio, "Quantum state engeneering
assisted by entanglement," Phys. Rev. A 67, 042104
(2003).

\bibitem{tomography1}G. M. D'Ariano and P. Lo Presti, "Quantum tomography for
measuring experimentally the matrix elements of an arbitrary
quantum operation," Phys. Rev. Lett. 86, 4195每4198 (2001).

\bibitem{tomography2} G. M. D'Ariano, P. Lo Presti, and M. G. A. Paris, "Using entanglement
improves the precision of quantum measurements," Phys.
Rev. Lett. 87, 270404-270407 (2001).

\bibitem{dolocalization} L. Heaney, A. Cabello, M. F. Santos, and V. Vedral, "Extreme
nonlocality with one photon," New J. Phys. 13, 053054每053065 (2011).
\bibitem{robust1} A. SenDe, U. Sen, M. Wie$\acute{s}$iak, D. Kaszlikowski, and M.
$\dot{Z}$kowski, "Multiqubit W states lead to stronger nonclassicality
than Greenberger-Horne-Zeilinger states," Phys. Rev. A 68,
062306 (2003).

\bibitem{robust2}  W. D$\ddot{u}$r, G. Vidal, and J. I. Cirac, "Three qubits can be entangled
in two inequivalent ways," Phys. Rev. A 62, 062314 (2000).


\bibitem{robust3}  R. Chaves and L. Davidovich, "Robustness of entanglement as a
resource," Phys. Rev. A 82, 052308 (2010).

\bibitem{SPE3} D. Gottesman, T. Jennewein, and S. Croke, "Longer-baseline telescopes using quantum repeaters," Phys. Rev. Lett. \textbf{109}, 070503 (2012).
\bibitem{distribution1} R. T. Thew, S. Tanzilli, W. Tittel, H. Zbinden, and N. Gisin, "Experimental investigation of the robustness of partially entangled qubits over 11 km," Phys. Rev. A \textbf{66}, 062304 (2002).

\bibitem{distribution2} I. Marcikic, H. de Riedmatten,W. Tittel, H. Zbinden, M. Legr$\acute{e}$,
and N. Gisin, "Distribution of time-bin entangled qubits over 50 km of optical fiber," Phys. Rev. Lett. \textbf{93}, 180502 (2004).

\bibitem{distribution3} T. Inagaki, N. Matsuda,O. Tadanaga, M. Asobe, and H. Takesue, "Entanglement distribution over 300 km of fiber,"
Opt. Express \textbf{21}, 23241-23249 (2013).

\bibitem{NLA1} T. C. Ralph and A. P. Lund, "Nondeterministic noiseless linear
amplification of quantum systems," in Proceedings of the 9th International Conference on Quantum Communication
Measurement and Computing, A. lvovsky, ed. (AIP, 2009),
pp. 155-160.

\bibitem{NLA2} G. Y. Xiang, T. C. Ralph, A. P. Lund, N. Walk, and G. J. Pryde,
"Heralded noiseless linear amplification and distillation of
entanglement," Nat. Photonics \textbf{4}, 316-319 (2010).

\bibitem{NLA3} N. Gisin, S. Pironio, and N. Sangouard, "Proposal for implementing
device-independent quantum key distribution based on a
heralded qubit amplifier," Phys. Rev. Lett. \textbf{105}, 070501 (2010).

\bibitem{NLA4}  M. Curty and T. Moroder, "Heralded-qubit amplifiers for practical
device-independent quantum key distribution," Phys. Rev. A
\textbf{84}, 010304(R) (2011).

\bibitem{NLA5} D. Pitkanen, X. Ma, R. Wickert, P. van Loock, and
N. L邦tkenhaus, "Efficient heralding of photonic qubits with applications
to device-independent quantum key distribution,"
Phys. Rev. A \textbf{84}, 022325 (2011).

\bibitem{NLA6}  C. I. Osorio, N. Bruno, N. Sangouard, H. Zbinden, N. Gisin, and
R. T. Thew, "Heralded photon amplification for quantum communication,"
Phys. Rev. A \textbf{86}, 023815 (2012).

 \bibitem{NLA7} S. Kocsis, G. Y. Xiang, T. C. Ralph, and G. J. Pryde, "Heralded noiseless amplification of a photon polarization qubit," Nat.
Phys. \textbf{9}, 23-28 (2012).

\bibitem{NLA8} S. L. Zhang, S. Yang, X. B. Zou, B. S. Shi, and G. C. Guo, "Protecting
single-photon entangled state from photon loss with noiseless linear amplification," Phys. Rev. A \textbf{86}, 034302 (2012)
.
\bibitem{NLA9} L. Zhou and Y. B. Sheng, "Distilling single-photon entanglement from photon loss
and decoherence," J. Opt. Soc. Am. B  \textbf{30}, 2737-2741 (2013).

\bibitem{NLA10} T. J. Wang, C. Cao, and C. Wang, "Linear-optical implementation of hyperdistillation from photon loss," Phys. Rev. A \textbf{89}, 052303 (2014).

\bibitem{NLA11} N. A. McMahon, A. P. Lund, and T. C. Ralph, "Optimal architecture for a nondeterministic noiseless linear amplifier," Phys. Rev. A \textbf{89}, 023846 (2014).

\bibitem{NLA12}  S. L. Zhang, Y. L. Dong, X. B. Zou, B. S. Shi, and G. C. Guo, "Continuous-variable-entanglement distillation with photon addition," Phys. Rev. A \textbf{88}, 032324 (2013).

\bibitem{NLA13} Y. B. Sheng, Y. Ou-Yang, L. Zhou, and L. Wang, "Protecting sing-photon multi-mode W state from
photon loss," Quantum Inf. Process. \textbf{13}, 1595-1605 (2014).

\bibitem{NLA14} J. Min$\acute{a}\breve{r}$, H. de Riedmatten, and N. Sangouard, "Quantum repeaters based on heralded qubit amplifiers," Phys. Rev. A \textbf{85}, 032313 (2012).

\bibitem{NLA15} Y. Ou-Yang, Z. F. Feng, L. Zhou, Y. B. Sheng, "Protecting single-photon entanglement with imperfect
single-photon source," Quantum Inf. Process. \textbf{14}, 635-651 (2015).

\bibitem{nonlinear} L. Zhou and Y. B. Sheng, "Recyclable amplification protocol for the
single-photon entangled state," Laser Phys. Lett. \textbf{12}, 045203 (2015).

\bibitem{timebin}  N. Bruno, V. Pini, A. Martin, B. Korzh,
 F. Bussi$\grave{e}$res, H. Zbinden, N. Gisin, and R. Thew,  "Heralded amplification of photonic qubits," Opt. Express \textbf{24}, 125-133 (2016)

 \bibitem{zhoutimebin} L. Zhou, and Y. B. Sheng, "The heralded amplification for the single-photon entanglement of the time-bin qubit" arXiv:1605.09480v1

\bibitem{photonefficiency} V. D'Auria, N. Lee, T. Amri, C. Fabre, and J. Laurat, "Quantum decoherence of single-photon counters," Phys. Rev. Lett. \textbf{107}, 050504 (2011)
\bibitem{photonefficiency1}  A. E. Lita, A. J. Miller, S. W. Nam,  "Counting near-infrared single-photons with 95\% efficiency," Opt. Express \textbf{16}, 3032-3040 (2008)


\end{thebibliography}
\end{document}